\documentclass[aps,twocolumn,showpacs]{revtex4}
\usepackage{graphicx}

\begin{document}

\title{Chaotic diffusion for particles moving in a time dependent potential 
well}

\author{$^1$Edson D.\ Leonel, $^1$C\'elia Mayumi Kuwana, $^1$Makoto Yoshida, 
$^2$Juliano Antonio de Oliveira}

\address{$^1$Universidade Estadual Paulista (UNESP) - Departamento de 
F\'isica\\ Av.24A, 1515 -- Bela Vista -- CEP: 13506-900 -- Rio Claro -- SP -- 
Brazil\\
$^2$Universidade Estadual Paulista (UNESP) - Campus de S\~ao Jo\~ao da 
Boa Vista\\ Av. Prof$^a$. Isette Corr\^ea Font\~ao, 505 -- CEP: 13876-750 -- 
S\~ao Jo\~ao da Boa Vista -- SP -- Brazil}

\date{\today} \widetext


\begin{abstract}
The chaotic diffusion for particles moving in a time dependent potential well 
is described by using two different procedures: (i) via direct evolution of 
the mapping describing the dynamics and ; (ii) by the solution of the diffusion 
equation. The dynamic of the diffusing particles is made by the use of a two 
dimensional, nonlinear area preserving map for the variables energy and time. 
The phase space of the system is mixed containing both chaos, periodic regions 
and invariant spanning curves limiting the diffusion of the chaotic particles. 
The chaotic evolution for an ensemble of particles is treated as random 
particles motion and hence described by the diffusion equation. The boundary 
conditions impose that the particles can not cross the invariant spanning 
curves, serving as upper boundary for the diffusion, nor the lowest energy 
domain that is the energy the particles escape from the time moving potential 
well. The diffusion coefficient is determined via the equation of the mapping 
while the analytical solution of the diffusion equation gives the probability 
to 
find a given particle with a certain energy at a specific time. The momenta of 
the probability describe qualitatively the behavior of the average energy 
obtained by numerical simulation, which is investigated either as a function of 
the time as well as some of the control parameters of the problem.
\end{abstract}

\maketitle

\section{Introduction}

Diffusive processes are observed in a wide variety of systems ranging from 
pollen diffusing from plant to plant \cite{paper1}, in medicine to the 
investigation of drugs delivered through the blood current \cite{paper2}, how 
diseases propagate in the air \cite{africa, malaria1, malaria2}, population 
movements in European prehistory \cite{culture}, pollution in atmosphere 
\cite{paper3} and in the oceans \cite{paper4}, percolation \cite{paper5} with 
the investigation of the chemical compounds being transported to the water 
table, memes in social media \cite{diego}, the influence of the clime in pests 
spreading \cite{pests} and many other applications from physical side 
\cite{paper6,paper7,paper8} turning the topic of interest to many 
\cite{paper9,paper10,paper11}.

In this paper we study the diffusion of chaotic orbits where particles are 
confined to move in a time dependent potential well \cite{paper12,paper13}. The 
physical motivation of such a potential \cite{paper14} comes from the 
interaction of photons with the array of atoms producing phonons in the chain 
leading to oscillations of the bottom part of the potential. The array is 
composed of infinitely many symmetric wells that can be modeled, due to the 
symmetry, by a single oscillating potential \cite{paper15}. The dynamic of each 
particle is made by the use of a two dimensional mapping given by energy of the 
particle and the corresponding time when it leaves the oscillating well. The 
oscillating part is characterized by three relevant control parameters, one 
gives the length of the well, other gives its depth and a third one gives the 
frequency of oscillation. The constant part of the potential has a parameter 
giving its length, which shall be considered symmetric with the time moving 
part. The phase space of the model is mixed and contains a set of periodic 
islands surrounded by a chaotic sea limited by a set of invariant spanning 
curves. Since the determinant of the Jacobian matrix is the unity, the 
particles moving in the chaotic sea can not penetrate the islands nor cross 
through the invariant spanning curve \cite{paper16} at the cost of violating 
the Liouville's \cite{paper6} theorem. The chaotic orbits diffuse along the 
chaotic sea and they behave similar to random walk particles at such a region. 
To describe their diffusion we solve the diffusion equation \cite{paper7} 
imposing that the flux of particles through the lowest energy invariant 
spanning curve is null as well as it is null in lower part of the phase space, 
where the particle reaches minimal energy to move outside of the time dependent 
part of the potential. The initial conditions are chosen all with a fixed 
energy 
$e_0$ and considering an ensemble of particles with such energy but with 
different phases of the moving potential. The solution of the diffusion 
equation \cite{paper7} gives the probability to find a given particle with a 
specific energy in a given time. From the knowledge of the probability, we 
obtain their momenta as functions of the time and compare with the numerical 
simulations. As we shall see, the qualitative agreement of the curves 
describing 
the diffusion is remarkable showing difference at the stationary state, a 
behavior obtained for long enough time. As we discuss along the text, such a 
difference is mostly connected to the shape of the phase space where islands 
are 
present surrounded by chaotic sea. As an assumption for the solution of the 
diffusion equation, such islands are not taken into account \cite{paper11} and 
their existence change the saturation of the curves for the diffusion at long 
enough time. We estimate the fraction that the islands occupy in the phase 
space and use it to made a first order correction in the saturation of the 
curves as an attempt to consider the same density of points in the phase space 
generated by the set of equations of the mapping and that used in the solution 
of the diffusion equation.

This paper is organized as follows. In Section \ref{sec2} we describe the 
equations of the mapping, construct the phase space and describe the estimation 
of the lowest energy invariant spanning curve. Section \ref{sec3} is devoted to 
discuss the results for the diffusion equation. Conclusion and final remarks 
are drawn in Section \ref{sec4}.

\section{The model, the mapping and some dynamical properties}
\label{sec2}

\begin{figure}[htb]
\centerline{\includegraphics[width=0.6\linewidth]{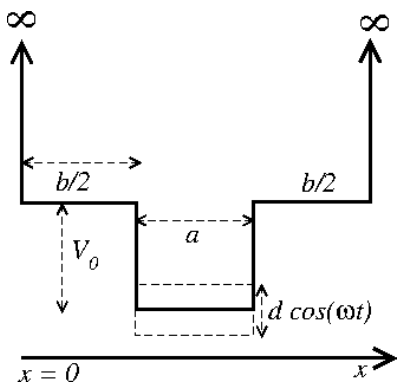}}
\caption{Sketch of the potential.} \label{Fig1}
\end{figure}

The dynamical model we consider consists of an ensemble of non interacting 
particles confined to move inside a potential well described by 
$V(x,t)=V_0(x)+V_1(x,t)$ 
where
$$
V_0(x)=\left\{\begin{array}{ll}
V_0~~~{\rm if}~~~0<x<{{b}\over{2}}~~ {\rm and}~~a+{{b}\over{2}}<x<a+b\\
~0~~~{\rm if}~~~{{b}\over{2}}\le x \le a+{{b}\over{2}}\\
\infty~~~{\rm if}~~~x\le 0~~{\rm and}~~x\ge a+b
\end{array}
\right.
$$
and $V_1(x,t)$ is written as
$$
V_1(x,t)=\left\{\begin{array}{ll}
~0~~~{\rm if}~~~x<{{b}\over{2}}~~ {\rm and}~~x>a+{{b}\over{2}}\\
d\cos(\omega t)~~~{\rm if}~~~{{b}\over{2}}\le x \le a+{{b}\over{2}}
\end{array}
\right..
$$
Figure \ref{Fig1} shows a plot of the potential.

\begin{figure}[t]
\centerline{\includegraphics[width=1.0\linewidth]{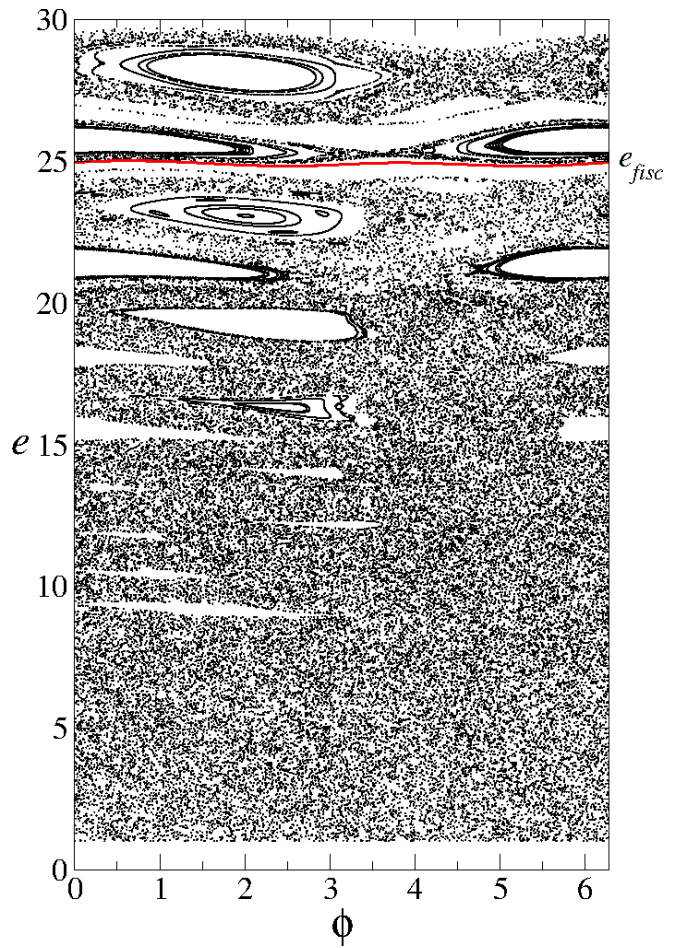}}
\caption{Plot of the phase space for mapping (\ref{eq1}) for the 
parameters: $r=1$, $\delta=0.5$ and $N_c=50$. The continuous curve identifies 
the position of the lowest energy invariant spanning curve, marking the upper 
limit of the chaotic sea.}
\label{Fig2}
\end{figure}

The dynamic of the particle is described by a two dimensional mapping written 
in terms of the energy and phase of the moving wall when it leaves the 
oscillating well. A careful discussion of the construction of the mapping can 
be 
found in Refs \cite{map1,map2,map3}. The expression of the mapping is given by
\begin{equation}
T:\left\{\begin{array}{ll}
e_{n+1}=e_n+\delta[\cos(\phi_n+i\Delta \phi_a)-\cos(\phi_n)],\\
\phi_{n+1}=[\phi_n+i\Delta\phi_a+\Delta\phi_b]~~{\rm mod (2\pi)},\\
\end{array}
\right.
\label{eq1}
\end{equation}
where $e_n=E_n/V_0$, 
$\Delta\phi_a={{2\pi N_c}\over{\sqrt{e_n-\delta\cos(\phi_n)}}}$ and 
$\Delta\phi_b={{2\pi N_cr}\over{\sqrt{e_{n+1}-1}}}$ with $r=b/a$, 
$\delta=d/V_0$ and $N_c={{\omega}\over{2\pi}}{{a}\over{\sqrt{2V_0/m}}}$. The 
parameter $N_c$ corresponds to the number of oscillations the potential well 
completes in the interval of time a particle travels the distance $a$ with 
kinetic energy $K=V_0$. Since it is proportional to $\omega$, increasing $N_c$ 
is equivalent to increase the driving frequency of the potential well. The 
variable $i$ corresponds to the smallest integer number, $i=1,2,3\ldots$ that 
satisfies the condition $e_n+\delta[\cos(\phi_n+i\Delta 
\phi_a)-\cos(\phi_n)]>1$ (see Ref. \cite{paper15,reflection1} for a discussion 
on the number of successive reflections in the well). Figure \ref{Fig2} shows a 
plot of the phase space for the mapping (\ref{eq1}) considering the control 
parameters $r=1$, $\delta=0.5$ and $N_c=50$.
We see a mixed structure of the phase space containing both periodic islands, 
a chaotic sea that is limited by a lowest energy invariant spanning curve. Such 
curve works as a barrier blocking the passage of particles through it confining 
the chaotic sea and defining an upper bound for the diffusion. The lower part 
is defined by the minimal energy the particle needs to escape the moving 
potential well, i.e., $e_{min}=1$. Therefore, the chaotic sea is limited to the 
range $e\in[e_{min},e_{fisc}]$. Along this paper we choose the parameter $r=1$ 
that characterizes the symmetric case. The two parameters we choose to vary are 
$\delta$ and $N_c$. The first parameter, $\delta\in[0,1]$ defines the amplitude 
of the oscillating well while the latter is proportional to the frequency of 
oscillation of the moving well, hence a parameter giving how fast the movement 
of the time dependent potential is.

\begin{figure}[t]
\centerline{\includegraphics[width=1.0\linewidth]{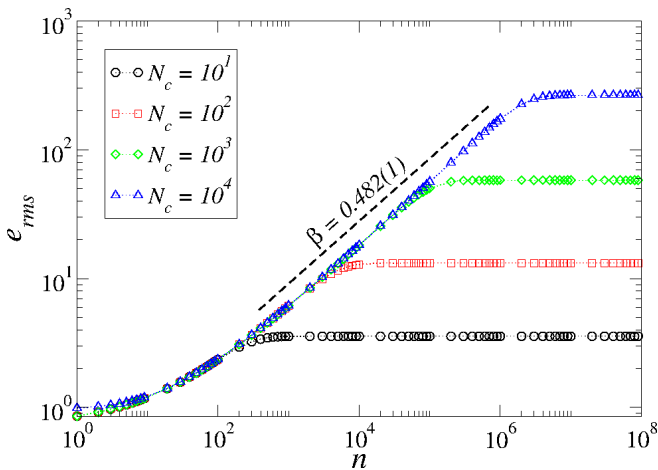}}
\caption{Plot of the curves of $e_{rms}~vs.~n$ for different values of 
$N_c\in[10^1,10^4]$ and $\delta=0.25$ for the symmetrical case $r=1$.}
\label{Fig3}
\end{figure}

\begin{figure}[t]
\centerline{\includegraphics[width=1.0\linewidth]{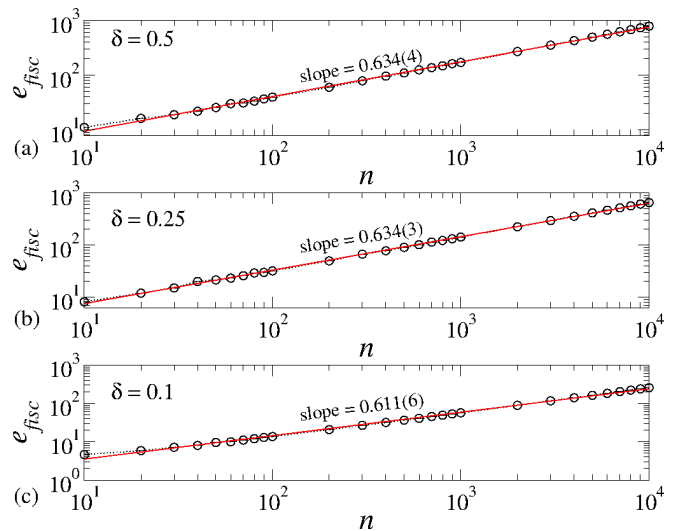}}
\caption{Plot of the average energy of the lowest energy invariant spanning 
curve. The parameters used were $r=1$, $N_c\in[10^1,10^4]$ and: 
(a) $\delta=0.5$, (b) $\delta=0.25$ and (c) $\delta=0.1$.}
\label{Fig4}
\end{figure}

The observable we are interested in investigate and that gives the 
characteristics of the diffusion is defined as
\begin{equation}
e_{rms}(n)=\sqrt{{{1}\over{M}}\sum_{i=1}^M \frac{1}{n} \sum_{j=1}^{n}e^2_{i,j}(n)},
\label{eq2}
\end{equation}
where $M$ gives the size of the ensemble of different initial phases
$\phi_0\in[0,2\pi]$ and $n$ is the number of iterations of the map. A typical behavior of the curves of $e_{rms}~vs.~n$ is 
shown in Fig. \ref{Fig3} for the parameters $N_c\in[10^1,10^4]$ and 
$\delta=0.25$ for the symmetrical case of $r=1$. 
We see from Fig. \ref{Fig3} that the curves of $e_{rms}$ start to growth in 
power law in $n$. A power law fitting of the regime of acceleration 
furnishes an exponent $\beta=0.482(1)$ which is very close to the exponent 
measured in normal diffusion for random walk particles. The saturation of the 
curves depends on the position of the lowest energy invariant spanning curve. 
Since the curves of $\lim_{n\rightarrow\infty}e_{rms}(n)$ depend on the size of 
the chaotic sea, the localization of the invariant spanning curves 
plays a major rule on the diffusion. Figure \ref{Fig4} shows a plot of the 
average energy of the lowest energy invariant spanning curve.

The results shown in Fig. \ref{Fig4} are in well agreement with the ones 
already known from the scaling theory \cite{paper15} where the formalism is 
based on three scaling hypotheses \cite{juliano} and from the localization of 
the invariant spanning curves \cite{spanning}. Starting with low initial energy, 
it gives the ensemble of particles the chance of experience the larger 
diffusion as possible. At this domain of energy, the behavior of $e_{rms}(n)$ is 
described by: (i) for short $n$, the curves exhibit diffusion such that 
$e_{rms}(n)\propto n^{\beta}$, hence the particles diffuse with slope 
$\beta\cong 0.5$. This numerical value gives evidence that the ensemble of 
chaotic particles are diffusing such as random particles. (ii) For large enough 
$n$, 
the particles visit a vast part of the phase space such that the curve of 
$\lim_{n\rightarrow\infty} e_{rms}(n)$ approaches a regime of saturation marked 
by $\lim_{n\rightarrow\infty} e_{rms}(n)\propto N_c^{\alpha}$. The numerical 
value obtained for such a behavior is from the order of $2/3$, hence in 
agreement with the description of the invariant spanning curves 
\cite{spanning}. The changeover from growth to the saturation is given by 
$n_x\propto N_c^z$. These scaling hypotheses allow to describe $e_{rms}$ as a 
homogeneous and generalized function yielding in the following scaling law 
$z=\alpha/\beta$. Therefore, $z\cong 4/3$. This investigation is 
phenomenological, so far. In the next section we provide a discussion that 
involves the solution of the diffusion equation \cite{paper7}. From such a 
solution, the momenta of the distribution are obtained and the critical 
exponents 
appear in an analytical way. As we shall see the existence of the islands 
affects the saturation of the curves by using the diffusion equation while 
compared with the direct simulations from the equations of the motion. We argue 
that the fraction of occupied area of chaos is a possible explanation for the 
difference although the qualitative description of the diffusion equation 
is remarkable good as compared to the one obtained from the numerical 
simulations of the dynamical equations.

\section{The diffusion equation}
\label{sec3}

Since the curves of the average squared energy are well described using scaling 
formalism, let us now go further and discuss the chaotic diffusion by using the 
diffusion equation. It is written as
\begin{equation}
{{\partial P(e,n)}\over{\partial n}}=D{{\partial^2P(e,n)}\over{\partial e^2}},
\label{eq3}
\end{equation}
where $P(e,n)$ gives the probability to find a particle with energy 
$e\in[e_{min},e_{fisc}]$ at a given time $n$. The diffusion coefficient $D$ is 
obtained from the first equation of mapping (\ref{eq1}) by using 
$D={{\overline{e^2}_{n+1}-\overline{e^2_n}}\over{2}}$. Doing the proper 
calculations we end up with the result that $D={{\delta^2}\over{2}}$.

To obtain an unique solution for Equation (\ref{eq3}) we impose the following 
boundary conditions
\begin{equation}
\left.{{\partial P}\over{\partial e}}\right|_{e=e_{fisc},e_{min}=1}=0,
\label{eq4}
\end{equation}
that guarantee no flux of particles through the invariant spanning curve at 
$e=e_{fisc}$ and also no flux of particle below the minimal energy $e_{min}=1$. 
At the initial time, all initial energies are located at $e=e_0$ with 
$e_{min}<e_0<e_{fisc}$ and that the probability must satisfy a delta function 
of 
the type
\begin{equation}
P(e,0)=\delta(e-e_0).
\label{eq5}
\end{equation}

We solved the Equation (\ref{eq3}) by the usual method of separation of 
variables. A solution that satisfies the diffusion equation is
\begin{equation}
P(e,n)=N_0e^{-c^2n}\left[\tilde{a}\cos\left({{ce}\over{\sqrt{D}}}\right)
+\tilde{b}\sin\left({{ce}\over{\sqrt{D}}}\right)\right],
\label{eq6}
\end{equation}
where $N_0$, $\tilde{a}$, $\tilde{b}$ and $c$ are constants to be achieved  
demanding that $P(e,n)$ must satisfy the boundary conditions. 
At $e=e_{fisc}$ we must have
\begin{equation}
{{\tilde{a}}\over{\tilde{b}}}={{\cos\left({{ce_{fisc}}\over{\sqrt{D}}}\right)}
\over{\sin\left({{ce_{fisc}}\over{\sqrt{D}}}\right)}},
\label{eq7}
\end{equation}
while at $e_{min}=1$ 
\begin{equation}
{{\tilde{a}}\over{\tilde{b}}}={{\cos\left({{c}\over{\sqrt{D}}}\right)}
\over{\sin\left({{c}\over{\sqrt{D}}}\right)}}.
\label{eq8}
\end{equation}
When matching the Equations (\ref{eq7}) and (\ref{eq8}), we end up with
\begin{equation}
c={{\sqrt{D}k\pi}\over{e_{fisc}-1}},
\label{eq9}
\end{equation}
where $k=1,2,3,\ldots$.

The case $k=0$ must be treated separated and leads to 
$P(e,n)=\tilde{N_0}(E_0e+f)$ that must also attend to the boundary conditions 
with $\tilde{N_0}$, $E_0$ and $f$ constants. Incorporating the eigenvalue as 
well as the boundary conditions for any $k$, the probability is written as
\begin{widetext}
\begin{equation}
P(e,n)=a_0+\sum_{k=1}^{\infty}
\left[a_k
\cos\left({{k\pi e}\over{e_{fisc}-1}}\right)+
b_k\sin\left({{k\pi e}\over{e_{fisc}-1}}\right)\right]
e^{{{-Dk^2\pi^2n}\over{(e_{fisc}-1)^2}}}.
\label{eq10}
\end{equation}
\end{widetext}

The next step is then to apply the initial condition at $n=0$. To do that we 
notice that the delta function can be written by its Fourier series 
representation{\cite{butikov}}
\begin{equation}
\delta(x)={{1}\over{2\pi}}+{{1}\over{\pi}}\sum_{k=1}^{\infty}\cos(kx).
\label{eq12}
\end{equation}
When we do the transformation $x={{\pi e}\over{e_{fisc}-1}}$, use the property 
that $\delta(ax)={{1}\over{a}}\delta(x)$ we obtain
\begin{widetext}
\begin{equation}
\delta(e-e_0)={{1}\over{2(e_{fisc}-1)}}+{{1}\over{(e_{fisc}-1)}}\sum_{k=1}^{
\infty }
\left[\cos\left({{k\pi e}\over{e_{fisc}-1}}\right)\cos\left({{k\pi 
e_0}\over{e_{fisc}-1}}\right)+
\sin\left({{k\pi e}\over{e_{fisc}-1}}\right)\sin\left({{k\pi 
e_0}\over{e_{fisc}-1}}\right)
\right].
\label{eq13}
\end{equation}
\end{widetext}

Comparing Equations (\ref{eq10}) and (\ref{eq13}) at $n=0$ we obtain that the 
coefficients are written as
\begin{eqnarray}
a_0&=&{{1}\over{2(e_{fisc}-1)}},\\
a_k&=&{{1}\over{(e_{fisc}-1)}}\cos\left({{k\pi 
e_0}\over{e_{fisc}-1}}\right),\\
b_k&=&{{1}\over{(e_{fisc}-1)}}\sin\left({{k\pi e_0}\over{e_{fisc}-1}}\right),
\end{eqnarray}
for $k=1,2,3,\ldots$. Using this set of coefficients, the probability given 
by Eq. (\ref{eq10}) is normalized in the range $e\in[e_{min},e_{fisc}]$.

The observable we want to investigate along the chaotic sea producing 
the chaotic diffusion is defined as 
$e_{rms}(n)=\sqrt{\overline{e^2}(n)}$, where
\begin{equation}
\overline{e^2}(n)=\int_{e=1}^{e_{fisc}}e^2P(e,n)de.
\label{eq14}
\end{equation}
The expression with all terms of $\overline{e^2}(n)$ is presented in the 
Appendix for the interested reader. From now we limit to consider that
\begin{equation}
\overline{e^2}(n)={{e^3_{fisc}-1}\over{6(e_{fisc}-1)}}+\sum_{k=1}^{\infty}{{e^{-
{{Dk^2\pi^2n}\over{(e_{fisc}-1)^2}}}} \over{(k\pi)^3}}[S_1+S_2+S_3+S_4].
\label{eq_dif_sol}
\end{equation}
The term $e_{fisc}$ is not obtained analytically, although numerical results 
for different values of $\delta$ is plotted in Figure \ref{Fig4}. 

We discuss the result given in Equation (\ref{eq14}). One notices that in 
the limit of $n\rightarrow\infty$, the exponential term approaches zero, 
lasting only the distribution for the stationary state, i.e.
\begin{equation}
\lim_{n\rightarrow\infty}e_{rms}(n)=\sqrt{{{e^3_{fisc}-1}\over{6(e_{fisc}-1)}}}.
\label{eq_satura}
\end{equation}
We notice that for $e_{fisc}\gg 1$, $\lim_{n\rightarrow\infty}e_{rms}(n)\cong 
e_{fisc}/\sqrt{6}$, which is in well agreement with the results shown in Figure 
\ref{Fig4} hence confirming the saturation exponent $\alpha\cong 2/3$.

Let us now discuss the evolution of $e_{rms}$ for short $n$. Expanding the 
exponential term of $\sum_{k=1}^{\infty}{{e^{-
{{Dk^2\pi^2n}\over{(e_{fisc}-1)^2}}}} \over{(k\pi)^3}}$ in Taylor series 
and considering only the first order approximation when assuming that the 
leading term of the summation is dominated by $k=1$, we obtain
\begin{equation}
e_{rms}(n)\cong \sqrt{{{e^3_{fisc}-1}\over{6(e_{fisc}-1)}}+{{S}\over{\pi^3}}-
{{SD}\over{\pi(e_{fisc}-1)^2}}n},
\end{equation}
where $S=S_1+S_2+S_3+S_4<0$ with $S_i$ for $i=1,2,3,4$ as show in the Appendix 
considering only the dominant term $k=1$. Therefore prior the saturation, the 
curves of $e_{rms}(n)\cong \sqrt{-{{SD}\over{\pi(e_{fisc}-1)^2}}n}$, hence 
$\beta=1/2$ as observed in the growing regime.

The regime marking the changeover from growth to the saturation can be 
estimated by doing 
$\sqrt{{{e^3_{fisc}-1}\over{6(e_{fisc}-1)}}+{{S}\over{\pi^3}}-
{{SD}\over{\pi(e_{fisc}-1)^2}}n}=\sqrt{{{e^3_{fisc}-1}\over{6(e_{fisc}-1)}}}$, 
leading to
\begin{equation}
n_x\propto {{(e_{fisc}-1)^2}\over{\pi^{2} D}}.
\end{equation}
Since $e_{fisc}\propto N_c^{\alpha}$, therefore for large $e_{fisc}$ we 
conclude that the critical exponent $z\cong 4/3$, as known in the literature 
\cite{juliano}.

\begin{figure}[t]
\centerline{\includegraphics[width=1.0\linewidth]{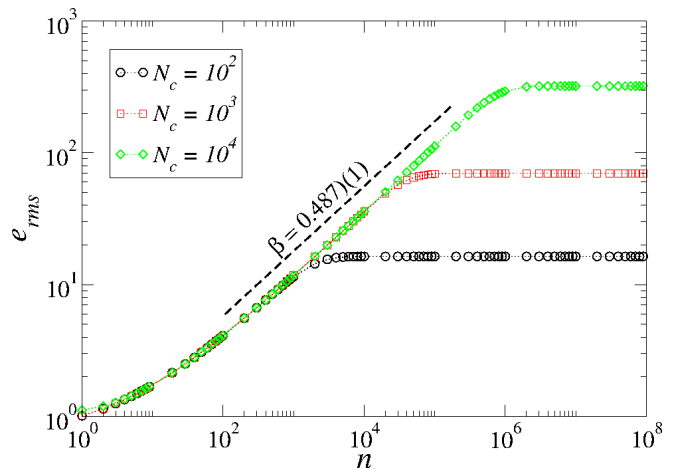}}
\caption{Plot of $e_{rms}(n)~vs.~n$ for different values of $N_c$, as 
labeled in the figure considering $r=1$ and $\delta=0.5$.}
\label{Fig5}
\end{figure}

The behavior of $e_{rms}(n)=\sqrt{\overline{e^2}(n)}$ obtained from Equation 
(\ref{eq_dif_sol}) for different values of 
$N_c$ is shown in Figure \ref{Fig5}.
We notice that the regime of growth is mostly marked by a slope of 
$\beta=0.487(1)\cong0.5$ when eventually the curves bend towards a regime of 
saturation as described by Eq. (\ref{eq_satura}).

\begin{figure}[t]
\centerline{\includegraphics[width=1.0\linewidth]{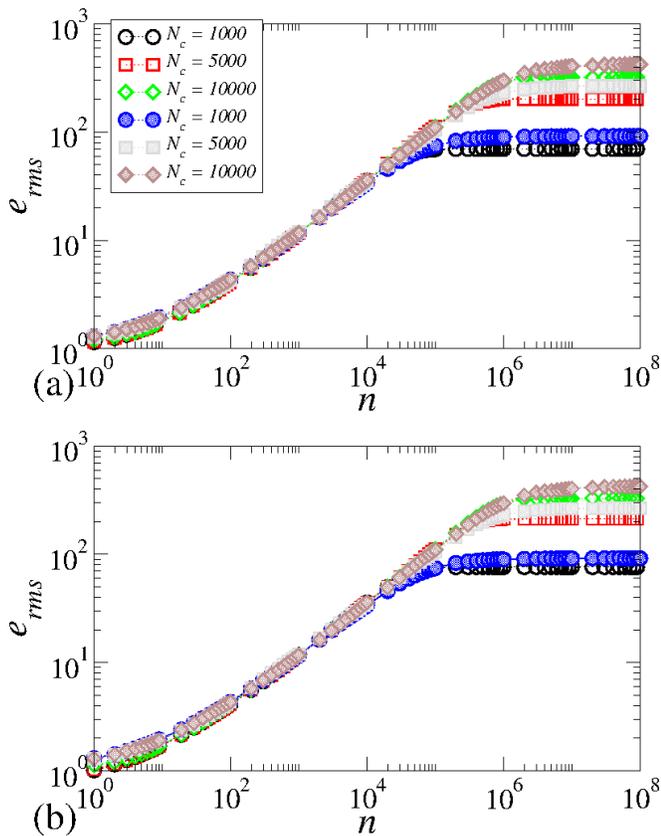}}
\caption{(a) Plot of $e_{rms}(n)~vs.~n$ for different values of $N_c$, as 
labeled in the figure considering $r=1$ and $\delta=0.5$. Filled symbols 
correspond to the numerical simulations while open symbols are the analytical 
results. (b) Same as shown in (a) but using the fraction of occupied phase 
space from chaotic dynamics.}
\label{Fig6}
\end{figure}

Let us now compare the analytical results with the numerical simulations. The 
better way of doing this is by plotting the curves of $e_{rms}(n)$ obtained 
from the different procedures in the same plot, as shown in Figure 
\ref{Fig6}(a).
The filled symbols correspond to the numerical simulations while open symbols 
are the analytical results. The agreement between the two procedures is 
remarkable good for short and intermediate values of $n$, either starting with 
$e_0\approx 1$ or larger values of $e_0$. The difference lies at the 
saturation. We now comment on such a difference. From the plot of the phase 
space, as seen in Figure \ref{Fig2}, one can see the existence of the 
periodicity islands. Since the mapping is area preserving, a chaotic orbit can 
not penetrate the islands. The orbit moves around it, sometimes getting sticky 
in such a region and eventually moves away, never getting inside. Because of 
the existence of the islands, the values of the saturation do not coincide with 
the half value of the chaotic window. However, at the conception of the 
boundary conditions in the diffusion equation, we assumed that the phase space 
is uniformly filled and that the existence of the periodic islands is not 
taken into account. This may be a possible reason why the curves of $e_{rms}$ 
constructed from the probability obtained from the diffusion equation saturate 
at lower values, as compared with the results obtained from the numerical 
simulations.

\begin{figure}[t]
\centerline{\includegraphics[width=1.0\linewidth]{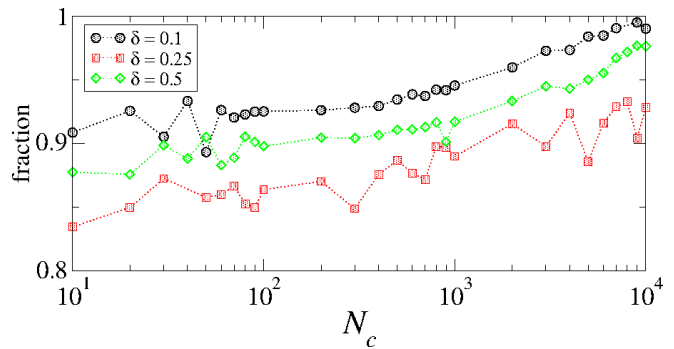}}
\caption{Plot of fraction of area filled by the chaotic sea in the chaotic 
region of the phase space limited at the range $e\in[e_{min},e_{fisc}]$. The 
parameters used were $r=1$ and $\delta=0.1$ (circle), $\delta=0.25$ (rectangle) 
and $\delta=0.5$ (losangle).}
\label{Fig7}
\end{figure}

To estimate the size of the chaotic sea in the allowed region of the phase 
space, we obtained the fraction of the area occupied by the chaotic sea for 
different values of $N_c$ and considering three different values of $\delta$. 
Figure \ref{Fig7} shows a plot
of the occupied area of the chaotic sea in the range $e\in[e_{min},e_{fisc}]$. 
Since the fraction is less than the unity, the density of points considered in 
the boundary conditions is different from the one assumed in the numerical 
simulations to that used in the solution of the diffusion equation. To have the 
same density of points in the phase space that takes 
into account the existence of the islands, the boundary conditions used in the 
diffusion equations would request a higher position of the invariant spanning 
curves. This difference in the density of points in the phase space from the 
one used in the solution of the diffusion equation is interpreted as being a 
possible reason for the curves obtained from the two different procedures 
saturate at distinct places. It is important to mention that such a difference 
was not observed in Ref. \cite{paper11} because the phase space considered 
there is symmetric with respect to the “action” axis in the positive and 
negative sides. The islands present in the positive side of $I$ in Ref. 
\cite{paper11} cancel the influence of the islands observed in the negative side 
of $I$ and such effect does not cause differences in the saturation between the 
theoretical point of view obtained from the solution of the diffusion equation 
and the numerical results, obtained directly from the dynamical equations. By 
using the fraction of the chaotic sea to correct the density of points in the 
phase space, the curves of $e_{rms}(n)$ obtained from the diffusion equation 
saturate closer to the ones obtained from numerical simulations, as seen in 
Fig. \ref{Fig6}(b).

\section{Conclusions and final remarks}
\label{sec4}

As a summary, we investigate the problem of chaotic diffusion for an ensemble 
of non interacting particles moving inside a time dependent potential well. The 
dynamic of each particle is described by an area preserving mapping for the 
dynamical variables energy and time when the particle leaves the moving well. 
The phase space is mixed with particles diffusing chaotically along a limited 
domain given by $e\in[e_{min},e_{fisc}]$. Starting with an initial condition 
$e_0\cong 1$, the curves of $e_{rms}$ growth to start with $n$ in a power law 
fashion of exponent $\beta\cong0.5$. This exponent allows to interpret chaotic 
diffusion similar to random walk particles. The saturation is marked by a power 
law $\lim_{n\rightarrow\infty}e_{rms}(n)=N_c^{\alpha}$ with $\alpha\cong 2/3$ 
in well agreement with the results known in the literature \cite{juliano}. The 
crossover marking the changeover from growth to the saturation is $z\cong4/3$, 
also in well agreement with the literature \cite{juliano}. Our original 
contribution on this topic comes from the analytical solution of the diffusion 
equation, that provides a function $P(e,n)$, that is a probability to obtain a 
given particle with energy $e$ at a specific time $n$. The boundary conditions 
used imply that the particles can not cross the invariant spanning curves at 
$e=e_{fisc}$ nor the lowest limit of the energy $e_{min}=1$ at the cost of 
violating the Liouville's theorem. The momenta of the distribution are obtained 
from $P(e,n)$ and compared with the numerical 
results. We notice a remarkable agreement of the two procedures at short and 
intermediate time but that the saturation for long enough time happens at 
different positions. The explanation is related to the existence of islands of 
stability in the phase space, which are taken into account direct numerical 
simulations and that are not considered in the solution of the diffusion 
equation. Although the quantitative results obtained from the two procedures 
differ from each other at large enough $n$, they show an incredible good 
agreement for short and intermediate time. The critical exponents obtained from 
the solution of the diffusion equation are not affected at large enough time, 
therefore even though the results are slightly different at large $n$, they 
validate the scaling theory of critical exponents obtained earlier in the 
literature \cite{paper15}. As perspective of the present approach we plan to 
investigate the chaotic diffusion in dissipative mappings including a 
dissipative version of the standard mapping and also time dependent billiards 
when particles experience inelastic collisions with the border of the billiard.

EDL thanks support from CNPq (301318/2019-0) and FAPESP (2019/14038-6), 
Brazilian agencies. CMK thanks to CAPES for support.

\section*{Appendix}

Here we present the expression of $\overline{e^2}(n)$. It is obtained from 
$\overline{e^2}(n)=\int_{e=1}^{e_{fisc}}e^2P(e,n)de$, leading to
\begin{equation}
\overline{e^2}(n)={{e^3_{fisc}-1}\over{6(e_{fisc}-1)}}+\sum_{k=1}^{\infty}{{e^{-
{{Dk^2\pi^2n}\over{(e_{fisc}-1)^2}}}} \over{(k\pi)^3}}[S_1+S_2+S_3+S_4]
\end{equation}
where the auxiliary terms are
\begin{widetext}
\begin{eqnarray}
S_1&=&\cos\left({{k\pi e_0}\over{e_{fisc}-1}}\right)
\left[\left(\pi^2k^2e^2_{fisc}-2(e_{fisc}-1)^2\right)\sin\left({{k\pi 
e_{fisc}}\over{ e_{fisc-1}}}\right)+2\pi(e_{fisc}-1)ke_{fisc}\cos\left({{k\pi 
e_{fisc}}\over{e_{fisc}-1}} \right)
\right],\nonumber\\
S_2&=&\sin\left({{k\pi e_0}\over{e_{fisc}-1}}\right)\left[
(2(e_{fisc}-1)^2-\pi^2k^2e^2_{fisc})\cos\left({{k\pi 
e_{fisc}}\over{e_{fisc}-1}}\right)
+2\pi(e_{fisc}-1)ke_{fisc}\sin\left({{k\pi e_{fisc}}\over{e_{fisc}-1}}\right)
\right],\nonumber\\
S_3&=&-\cos\left({{k\pi e_0}\over{e_{fisc}-1}}\right)
\left[\left(\pi^2k^2-2(e_{fisc}-1)^2\right)\sin\left({{k\pi 
}\over{e_{fisc-1}}}\right)+2\pi(e_{fisc}-1)k\cos\left({{k\pi 
}\over{e_{fisc}-1}}\right)\right],\nonumber\\
S_4&=&-\sin\left({{k\pi e_0}\over{e_{fisc}-1}}\right)\left[
(2(e_{fisc}-1)^2-\pi^2k^2)\cos\left({{k\pi}\over{e_{fisc}-1}}\right)
+2\pi(e_{fisc}-1)k\sin\left({{k\pi}\over{e_{fisc}-1}}\right)
\right].\nonumber
\end{eqnarray}
\end{widetext}


\begin{thebibliography}{9}

\bibitem{paper1} W. F. Morris, Ecology {\bf 74}, 493 (1993).

\bibitem{paper2} K. Murase, S. Tanada, H. Mogami, M. Kawamura, M. Miyagawa, 
M. Yamada, H. Higashiro, A. Lio, K. Hamamoto, Medical Physics {\bf 19}, 70 (1990).

\bibitem{africa} S. A. El-Kafrawy et al, The Lancet Planetary Health {\bf 3}, e521 (2019).

\bibitem{malaria1} Z. Xu, Y. Zhang, IMA J. of Applied Mathematics {\bf 80}, 1124 
(2015).

\bibitem{malaria2} Y. Lou, X. Q. Zhao, J. Math. Biol. {\bf 62}, 543 (2011).

\bibitem{culture} E. E. Harris, Evolutionary Anthropology {\bf 26}, 228 (2017).

\bibitem{paper3} D. Popp, Journal of Environmental Economics and Management {\bf 51}, 46 (2006).

\bibitem{paper4} R. V. Ozmidov, Diffusion of contaminants in the ocean, 
Springer (1990).

\bibitem{paper5} C. Hagedorn, E. L. Mc Coy, T. M. Rahe, J. Environ. Qual. {\bf 10}, 1 
(1981).

\bibitem{diego}X. Qiu, D. F. M. Oliveira, A. S. Shirazi, A. Flammini, F. 
Menczer, Nature Human Behaviour {\bf 1}, 0132 (2017). 

\bibitem{pests} W. S. Jo, H. Y. Kim, B. J. Kim, J. Korean Phys. Soc. {\bf 70}, 108 (2017). 

\bibitem{paper6} F. Reif, Fundamentals of statistical and thermal
physics, New York: McGraw-Hill (1965).

\bibitem{paper7} V. Balakrishnan, Elements of nonequilibrium statistical
mechanics, Ane Books India, New Delhi (2008).

\bibitem{paper8} R. K. Patria, Statistical Mechanics, Elsevier (2008).

\bibitem{paper9} E. D. Leonel, J. Penalva, R. M. N. Teixeira, R. N. Costa 
Filho, M. R. Silva, J. A. Oliveira, Physics Letters A {\bf 379}, 1808 (2015).

\bibitem{paper10} A. L. P. Livorati et al, Commun. Nonlinear Sci. Numer. Simulat. {\bf 55}, 225 (2018).

\bibitem{paper11} E. D. Leonel, C. M. Kuwana, J. Stat. Phys. {\bf 170}, 69 (2018).

\bibitem{paper12} J. L. Mateos, J. V. Jos\'e, Physica A {\bf 257}, 434 (1998).

\bibitem{paper13} J. L. Mateos, Phys. Lett. A {\bf 256}, 113 (1999).

\bibitem{paper14} G. A. Luna-Acosta, G. Orellana-Rivadeneyra, A. Mendoza-
Galv\'an, C. Jung, Chaos, Solitons Fractals {\bf 12}, 349 (2001).

\bibitem{paper15} D. R. Costa, M. R. Silva, J. A. Oliveira, E. D. Leonel, 
Physica A {\bf 391}, 3607 (2012).

\bibitem{paper16} A. J. Lichtenberg, M. A. Lieberman, Regular and chaotic
dynamics  (Appl. Math. Sci.) 38, Springer Verlag, New York (1992).

\bibitem{map1} E. D. Leonel, P. V. E. McClintock, Phys. Rev. E {\bf 70},
016214 (2004).

\bibitem{map2} E. D. Leonel, P. V. E. McClintock, Chaos {\bf 15}, 033701 
(2005).

\bibitem{map3} D. R. da Costa, C. P. Dettmann, E. D. Leonel, Phys. Rev. E 
{\bf 83}, 066211 (2011).

\bibitem{reflection1} E. D. Leonel, J. K. L. Silva, Physica A {\bf 323}, 181 
(2003).

\bibitem{juliano} J. A. de Oliveira, R. A. Biz\~ao, E. D. Leonel, Phys. Rev. E {\bf 81}, 
046212 (2010).

\bibitem{spanning} E. D. Leonel, J. A. de Oliveira, F. Saif, J. Phys. A {\bf 44}, 
302001 (2011).

\bibitem{butikov} E. Butkov, Mathematical Physics, Addison-Wesley Pub. Co., 
(1968)
\end{thebibliography}
\end{document}